# Comparison between two numerical schemes to study the spectra of charmed quarkonium

A. M. Yasser [1], G. S. Hassan[2], Samah K. Elshamndy [3, 4], M. S. Ali[2, 5]

1  Physics Department, Faculty of Science at Qena, South Valley University, Egypt.

2  Physics Department, Faculty of Science, Assiut University, Egypt.

3 University of Jouf, Faculty of Science and Art, Department of Physics, Post Office 207, Al-Jouf, KSA.

4 Radiotherapy and Nuclear Medicine Department, Faculty of Medicine, Sohag University, Egypt.

5 African Institute for Mathematical Sciences, B. P. 1418 Mbour, Thies Region, Senegal.

**Abstract**

Two numerical methods are developed to reduce the solution of the radial Schrödinger equation for proposed heavy quark-antiquark interactions, into the solution of the eigenvalue problem for the infinite system of tridiagonal matrices. Our perspective is a numerical approach relies on finding the proper numerical method to investigate the static properties of heavy quarkonia-mesons, such as spectrum, radius … etc., with implantation of both the nonrelativistic quark model and the ingredients of the quantum chromodynamics (QCD) theory. The application of these proposed schemes resulted in mass spectra of charmed-quarkonium (charmonium) multiplets, which are compared with the experimental published profiles of Particle Data Group (PDG). Besides, the normalized radial wave-functions of the charmonium various bound states are represented. The convergence of each numerical recipe versus the iteration number N and the radial distance is investigated through this work. Although it was observed that our numerical treatments are reliable to study charmed Quarkonium bound states profile, we found that one of these proposed techniques is favored over the other in terms of high precision comparisons with experiments and convergence analysis.

**Keywords:** Finite Difference method, Matrix Numerov's method, Quarkonium, Charmonium, Heavy mesons, Radial Schrödinger equation, Numerical analysis, Convergence analysis.

## 1. Introduction

Quarkonium is a term that refers to heavy quark-antiquark constituents that build up heavy mesons. November revolution, 1974, led to renewed interest for heavy mesons investigation, when a resonance foundation was discovered independently at Brookhaven National

Laboratory (BNL) {1} and at Stanford Linear Accelerator Center (SLAC) {2}. The new breakthrough was the first discovered bound state of charmed Quarkonium (charmonium), which contains charmed quark and its antiquark. Thus, it is commonly denoted as $c\bar{c}$ {3} {4} {5}.

The main motivation beyond the investigation of the mesonic world in the quark model framework is raised to be on the surface to understand the quark model applicability, and to generate possible improvements. Additionally, charmonium investigations were hoped to play the similar role for configuring hadronic colored-dynamics as the hydrogen atom studies played in understanding the atomic physics {6}. Currently, it is essential to study the properties of the deconfined state of nuclear matter, the Quark-Gluon Plasma (QGP), produced in ultra-relativistic heavy ion collisions.

Our perspective relies upon applying two developed numerical recipes of the solution of the Schrödinger equation to the calculation of the mass spectra of heavy charmed quarkonium consisting of heavy charmed quark and antiquark with implementation of the non-relativistic quark model. The associated assumption of the quark model in the nonrelativistic range, which is widely used for the heavy hadronic description, is the application of the nonrelativistic approximation. The efficiency of applying this approximation goes better with the increasing of the constituent quark masses.

Previous calculations in the framework of the relativistic quark model emphasized that the implementation of the nonrelativistic approximation provides a satisfying description of the static properties of heavy quarkonia, such as spectrum, radius, etc. On the other hand, the dynamical properties of heavy mesons need to apply relativistic corrections as suggested by W. Lucha et al {7}.

Recently, the experimental observations of charmonium-related multiplets have swiftly overstepped the ability of theoretical treatments to establish a consistent framework might fit their properties. From this perspective, the role of theorists comes to model the spectra of charmonium or to improve the previous treatments, e.g. computationally by using modern effective numerical methods, to be endorsed.

Many attempts to study mesonic sectors and their corresponding transitions, for instance the investigations of Kwong et al {8}{9}, Godfrey and Rosner et al {10}{11}, J. Segovia et al {12}{13}, and Kumar et al {14}. As a matter of fact, several numerical schemes are developed via these previous studies to study a broad variety of the hadronic sectors in both frameworks of quark model, relativistic and non-relativistic. Although these numerical treatments have been inferred a reliable applicability, extreme complexity could not be avoided. On the other hand, our approach relies on finding numerical methods to be much easier and faster computationally without losing high precision in term with comparisons with the experimental published data and theoretical results yielded via some of these models. Actually, this work is an essential step of a sequence of attempts proposed previously by our group seeking to achieve this aim {15}{16}{17}{18}.

Lattice calculations were the preliminary development of the gauge theory, which declared in 1974, it is a lattice formulation technique proposed for simple systems. Lattice calculations possess computational complexity, as it is extremely expensive {19}{20}. Its applicability is reduced recently with most known systems, as the complexity is increased with complicated systems. Many attempts to improve these schemes for complex systems are encountered by failure.

On the other hand, Fourier grid Hamiltonian techniques showed good applicability for complex systems in both relativistic and non-relativistic quark models {21}{22}{23}. The second crop is additional numerical treatments that have been developed in the previously mentioned quark models, such as Shooting method {24}, and Numerov's techniques {25}. For some cases, these numerical techniques showed tremendous computer time, which is aimed to be avoided in our numerical developments, although high accuracy is achieved. From this standpoint, our perspective is released to be on the surface. In our approach, two numerical developments are considered to be in comparisons with the published experimental data and also with published theoretical data, seeking to find the optimal technique to study the hadronic worlds.

According to the previous investigations, numerous interactions are associated with complex three-quarks systems by which the corresponding expression is not obtainable yet {26}. To avoid this dilemma, simplicity dictates to consider only the sum of two-body interactions, which is a good approximation for three-body interactions {27}. It is worth noting that this idea becomes much poorer for multi-quark systems as the number of quark increases, the number of forces increases correspondingly.

Genuinely, we hope to find an endorsed theoretical framework for investigation mesonic systems might its applicability extended to be involved in the baryonic sectors, additionally giving the gross features of the mesonic worlds. In view of the fact that is declared by R. Vinh Mau, C. Semay et al {28}, many-body forces do not play a major role in characterizing the hadron-hadron interactions. According to this standpoint of view, our attempt might give an incremental contribution to recognizing the hadronic sectors.

In general, finding the optimal solution of the spectral problem, the Schrödinger and Schrödinger-like equations with spherically symmetrical potentials is one of the essential problems for both atomic and hadronic spectroscopy. From this standpoint of view, the proposed approach here is appropriate computationally for analytic solutions and for numerical computations for Schrodinger equation and semi equations, such as Klein–Gordon–Fock equation and Dirac equation, for any arbitrary Hamiltonian. Thus, both methods allow determining the spectrum of the Schrodinger-like relativistic or non-relativistic equations. Thus, this makes our perspective considered as an incremental contribution not only for high-energy physics but also for other related branches, such as atomic physics, quantum mechanics, etc.

This paper is organized as follows, the second section presents the Hamiltonian model, which is associated with potential phenomenological terms that describes the different interactions between heavy charm and anticharm. While the third part demonstrates the proposed numerical methods used to reduce the solution of the radial Schrodinger equation for the proposed potential to the solution of the eigenvalue problem for the infinite system of tridiagonal matrices.

Theoretical mass spectra of both low-lying and high charmed quarkonia multiplets and the associated radial wavefunctions are illustrated in the fourth sections. Numerical analysis for the proposed numerical techniques is discussed with illustration for convergence curves versus iteration numbers N and radial distances in the fifth section. Ultimately, some concluding remarks are presented in the last section.

## 2. Heavy Quarkonia Governed Hamiltonian

The November revolution, in 1974, represented the charmed quarkonium system as prototypical to positronium, which is an exotic atom consisting of an electron and its antimatter the positron, of mesonic spectroscopy {29} {30} {31}. This attracted theorists' attention to find the correlation between the charmonium multiplets spectra and those of the positronium exotic atom, which seems roughly perfect {26} {5} {32}.

The observed similarities discovered after between charmed quarkonium system and the positronium exotic atom leads theorists to refine hydrogen or positronium-like potential models can be used to describe the dynamics of Quarkonia systems with the implementation of the ingredients of QCD theory as well.

The appropriate governed Hamiltonian for description of heavy charmed quarkonium dynamics, in the non-relativistic quark model, is prescribed of two parts: kinetic energy and potential energy, which takes into account the various phenomenological interactions between the heavy charm and anti-charm quarks.

$$\widehat{H}_{nr} = \widehat{T}_{nr} + \widehat{V}_{nr}$$

Relativistic corrections, which are derived from the QCD theory, are being applied to our non-relativistic Hamiltonian model to increase its applicability to describe dynamical properties of charmed quarkonia needed to apply these relativistic corrections, such as the decay of its multiplets. In spite of this, we restricted our attention to calculating only its mass spectra. The associated Kinetic energy of charmed quarkonium, according to C. Semay et. al. proposed Hamiltonian model, is given by {6}:

$$\hat{T}_{nr} = m_Q + m_{\underline{Q}} + \frac{\hat{P}^2}{\mu}$$

where $m_Q$ and $m_{\underline{Q}}$ are the masses of charmonium constituents, $\hat{P}$ is the system relative momentum and µ refers to the reduced mass of the system. As it is obvious, the K.E expression is adjusted with the masses of the charmed constituents. The reason beyond this is to make the calculations of heavy mesons mass spectra comparable with the observed experimental discoveries. This perspective is proposed by The Godfrey and Isgur (GI) model in 1985 and developed in C. Semay and B. Silvester-Brac (CB) treatments to investigate the properties of various mesonic family members. This approach is emphasized to be a reliable correction for the non-relativistic K.E term for heavy mesons {33}{6}.

Obtaining the desired description for mesonic mass spectrum relies on using appropriate potential models. The potential, which describes the interaction dynamics, for such a system is usually assumed to obey the confining type. For instance, Cornell potential is a good example of this important fact, which accounts for asymptotic freedom in QCD at small distances and the linear rise of the potential with the increasing distance between charm and anti-charm quarks. This type of potential model consists of two phenomenological terms. The first is responsible for the Coulomb-like interaction, which is dominating the dynamics at small radial distances. On the other hand, the second corresponds to the string-like interaction that leads to the confinement {17}.

Actually, spin-dependent interactions, which include both spin-spin interaction and spin-orbit interaction, should not be avoided in our approach. For instance, spin-spin interaction accounts the Gaussian smeared contact hyperfine interaction in the zeroth-order potential according to Oak Ridge National Laboratory study {34} [Assiut thesis]. This term of the potential model is associated with one-gluon exchange forces. The remaining terms of the spin-dependent part in our potential model are called phenomenologically mass shifts. These terms are obeying to the leading-order of the perturbation theory. These remaining expressions include one-gluon exchange spin-orbit term, tensor interaction part, and longer ranged inverted spin-orbit term. The last one arises from the assumed Lorentz scalar confinement interaction. Ultimately, the potential energy of the charmed quarkonium can be described by the previous mentioned phenomenological terms as following:

$$V_3(r) = \frac{-4}{3}\frac{\alpha_s}{r} + br + \frac{4\alpha_s}{3m_Q m_{\underline{Q}}}\frac{8\pi}{3}\delta(\vec{r})(\vec{S}_1.\vec{S}_2) + \frac{l(l+1)}{2\mu r^2} + \frac{1}{m_Q m_{\underline{Q}}}\left(\frac{2\alpha_s}{r^3} - \frac{b}{2r}\right)\vec{L}.\vec{S} + \frac{4\alpha_s}{r^3}T \quad 2.1$$

It worth to be mentioned that the spin-spin operator is considered to be diagonal in $|J, l, s>$ basis, which is associated with these matrix elements:

$$<\vec{L}.\vec{S}> = \frac{[J(J+1) - l(l+1) - S(s+1)]}{2}$$

The term of equation 2.1 involved the tensor operator T that can be described by non-vanishing diagonal matrix elements; only for $l > 0$; which refers to spin-triplet states, can be represented as:

$$T = \begin{cases} \frac{-l}{6(2l+3)}, & J = l+1 \\ \frac{1}{6}, & J = l \\ \frac{-l+1}{6(2l-1)}, & J = l-1 \end{cases}$$

While, the non-relativistic expression of kinetic energy is given by [4]:

$$T_{non-relativis} = T_{kinetic-energ} + m_1 + m_2$$

## 3. The numerical methods

In this study, the solution of the radial Schrödinger equation has been reduced to the solution of the eigenvalue problem for the infinite system of tri-diagonal matrices, which called the matrix method and Numerov's matrix method respectively. The matrix method will be discussed firstly.

### 3.1 Tri-diagonal matrix method

Various attempts have proposed numerical schemes to solve the time-independent Schrödinger equations for any spherical symmetric potential models {5} {32} {35}. In this section, we propose a type of tri-diagonalization scheme to reduce the solution of the radial Schrödinger equation into an eigenvalue problem depending on reducing the two-body problem of charm-anticharm system into one-body problem with reduced mass $\mu = m_c + m_{\underline{c}}/m_c m_{\underline{c}}$ at which the associated phenomenological dispersion distance from a particular point in the non-relativistic model is denoted by r. Thus, the associated Hamiltonian of this heavy-quarks configuration is spherically symmetric by which the radial Schrödinger equation is suitable to investigate its spectroscopy. The radial Schrödinger equation for heavy mesons systems can be represented as:

$$\frac{-1}{2\mu}\frac{\partial^2}{\partial r^2}U(r) + \left(\frac{1}{2\mu r^2}l(l+1) + [V(r) - E]\right)U(r) = 0 \qquad 3.1.1$$

where $U(r)$ is the associated radial wavefuction. Our perspective relies upon reducing this equation through diagonalization into tri-diagonal matrix form to be solved as an eigenvalue problem through which the phenomenological mass-spectra for charmed-Quarkonium multiplets can be obtained from the retrieved eigenvalues and also the normalized wavefunctions of these multiplets can be gotten via the crossponding eigenvectors.

For the sake of the simplicity, the diagonalization of this system of equation is performed through developing the finite difference approximation of the radial second derivative of the associated wavefuction of charmed-Quarkonium into a tri-diagonal matrix form. It worth to be mentioned that the finite difference quotient of the second derivative can be represented as:

$$\frac{d^2 U(r)}{dr^2} = \frac{U_{i+1} - 2U_i + U_{i-1}}{d^2} + O(\Delta r^3)$$

where d refers to the uniform mesh spacing of the used computational grid at which ith points to the sites through the computational grid that corresponds to the phenomenological radial position of our body in the system. Consequently, i=1,2,3,….(N-1), where N refers to the number of steps over the computational grid. The computational mesh spacing is phenomenologically obtained from the maximum and minimum radial distances $R_{max}$ and $R_{min}$ respectively as following:

$$d = \frac{R_{max} - R_{min}}{N}$$

Thus, any arbitrary position of our body can be defined through the following:

$$r_i = R_{min} + id$$

Consequently, the radial Schrödinger equations can be reduced into this form over a discrete space $r_i$ to be:

$$C_2 U_{i+1} + \left(C_1 + V(r_i) + \frac{1}{2\mu}\frac{l(l+1)}{r_i}\right)U_i + C_2 U_{i-1} = E U_i \qquad 3.1.2$$

Where $C_1 = \frac{1}{\mu d^2}$ and $C_2 = \frac{-1}{2\mu d^2}$.

Additionally, the previous equation can be simplified by considering:

$$h(i) = C_1 + V(r_i) + \frac{1}{2\mu}\frac{l(l+1)}{r_i^2}$$

Eventually, the tri-diagonal algorithm of the system of equations is expressed by:

$$C_2 U_{i+1} + h(i)U_i + C_2 U_{i-1} = E U_i \qquad 3.1.3$$

Consequently, the tridiagonal system of equations can be represented into a tri-diagonal matrix form of a computational grid $(N-1) \times (N-1)$ as following:

$$(h(1)\ C_2\ 0\ 0\ \ldots\ 0\ C_2\ h(2)\ C_2\ 0\ \ldots\ \ldots\ \ldots\ \ldots\ \ldots\ \ldots\ 0\ \ldots\ \ldots\ \ldots\ h(N-2)\ C_2\ 0\ \ldots\ \ldots\ \ldots\ C_2\ h(N-1))(U_1\ U_2\ \ldots\ \ldots\ U_{N-1}) = E(U_1\ U_2\ \ldots\ \ldots\ U_{N-1}) \qquad 3.1.4$$

Through the previous scheme, any two-body bound state problem can be reduced to be solved as an eigenvalue problem over all hadronic sectors. Additionally, it might be extended to be used for atomic and sub-atomic scales, which is considered as an incremental contribution for all theoretical physics fields.

### 3.2 Tri-diagonal matrix Numerov's method

As mentioned previously, tri-diagonal matrix method relies upon the finite difference scheme to approximate the solution of the Hamiltonian of the radial Schrödinger equation. Thus, the associated truncation error of the diagonalization using matrix method is obtained to be with order 3. Consequently, it seems to be useful to investigate new diagonalization ways that associated with high order local truncation errors.

The new diagonalization that used for reducing the solution of the radial Schrödinger equation into an eigenvalue problem relies on the well-known Numerov's method. Indeed, this numerical recipe is a widely known to approximate the solution of the ordinary differentiation equations (ODE) in form $U''(x) = f(x)U(x)$ {36}.

The phenomenological time independent Schrödinger equation for quark-antiquark configurations can be written in that form:

$$f(r) = 2\mu(E - V(r)) - \frac{l(l-1)}{r^2}$$

The previous expression can be diagonalized with Numerov's scheme using built-in tri-diagonal matrix routines [ref to the second paper] by which the resulting eigenvectors are used to generate the superposition of the basis states to represent the corresponding spatial wavefunctions of charmonium multiplets.

According to our numerical approach, the differential operator of the kinetic energy is represented on a computational grid with sites $r_i$ evenly separated with a specific computational distance $d$ in a straightforward manner. This manner is associated with high-accuracy solutions using the tri-diagonalization algorithms. Additionally, the operator of the potential energy is merely diagonalized into a tri-diagonal matrix of the charmonium potential energy assessed at each computational point.

The computational space is obtained from the maximum and the minimum values of the phenomenological radial distances of charmonium configuration as following:

$$d = \frac{r_{max} - r_{min}}{N}$$

where N is the number of the computational steps all over the grid. It worth to be mentioned, the i*th* points on the grid, which corresponds to the radial positions $r_i$, are determined by $i = 1, 2, \ldots, (N-1)$.

The integration formula of Numerov's scheme, over even-points computational grid, is represented by the following:

$$\psi_{i+1} = \frac{\left(2 - \frac{5d^2}{6} f_i\right)\psi_i - (1 + \frac{d^2}{12} f_{i-1})\psi_{i-1}}{(1 + \frac{d^2}{12} f_{i+1})} + O(\Delta r^6) \qquad 3.2.1$$

where $f_i$ is obtained from the previous expression of the time independent radial Schrödinger equation with considering $V_i = V(r_i)$ and $\psi_i = \psi(r_i)$ to get the following algorithm expression:

$$\frac{-1}{2\mu} \frac{(\psi_{i-1} - 2\psi_i + \psi_{i+1})}{d^2} + \frac{V_{i-1}\psi_{i-1} + 10 V_i \psi_i}{12} = E \frac{\psi_{i-1} + 10\psi_i + \psi_{i+1}}{12} \qquad 3.2.2$$

The previous expression can be rearranged into a tri-diagonalized matrix form to obtain the matrix Numerov's algorithm as following:

$$\frac{-1}{2\mu}(B^{-1}A)\psi + V\psi = E\psi \qquad 3.2.3$$

where $A = \frac{I_{-1} - 2I_0 + I_1}{d^2}$, $B = \frac{I_{-1} - 10 I_0 + I_1}{d^2}$, and $V = diag(\ldots V_{i-1}, V_i, V_{i+1} \ldots)$. It worth to be mentioned that $I_p$ matrix is a matrix of 1s along the p*th* diagonal entries, and zeros elsewhere.

Through this diagonalization that is associated with a 6-order local truncation error, we can reduce an arbitrary two-body problem, not only in mesonic sectors but it might be extended to involve all hadronic sectors, into an eigenvalue problem.

In the next step, we will represent the results of the charmed-Quarkonium profiles obtained by the previously elaborated tri-diagonal matrix methods with statistical and numerical investigations in terms of comparison to find the favored numerical scheme that is suitable to study the spectroscopy of the charmed-Quarkonium.

## 4. Results and discussion

Through this section of our study, the theoretical results, based on our numerical recipes, will be examined to be sure that the predicted masses do not blowup to infinity and the solutions of the reduced tri-diagonal eigenvalue matters are physical.

Optimization is involved our work through minimizing the statistical function $\chi^2$ to emphasize the endorsement between the theoretical predictions and experiments. All chosen experimental masses of charmed-Quarkonium multiplets in such comparisons are obtained from the review of Particle Data Group (PDG) {37} {38}. It worth to be mentioned, the statistical function is defined via:

$$\chi^2 = \frac{1}{n}\sum_{i=1}^{n} \frac{(M_i^{exp} - M_i^{theo})^2}{\Delta_i}$$

The summation in the previous expression runs over selected masses n of charmed-Quarkonium multiplets that possess discovered masses by which $M_i^{exp}$ refers to the observed masses of Charmonium multiplets, while $M_i^{theo}$ are the corresponding theoretical masses depending on both the phenomenological potential model selected and the used numerical scheme. Additionally, the $\Delta_i$ quantity points to the experimental uncertainty on the corresponding mass. Intuitively, $\Delta_i$ should be one.

Through this section, the normalized radial wavefunctions of some samples of charmonium states will be illustrated with considering to the modern nomenclature system of heavy mesons spectroscopic notation.

Illustration in this section involves the stability curves for samples of lower states of charmonia, which own L=0 and S=0 or 1, and also some samples of higher states that possess L≥1and S=0 or 1 for chosen numerical recipes. The stability curves display the conversion of the inverse of predicted masses of charmonia versus both the computational iteration number (N) and the phenomenological radial distance between charm and anti-charm in our system. This investigation is important to examine the reliability of our numerical schemes to be sure that these methods doesn't blowup to infinity, and ultimately to investigate the cheap method computationally.

In our approach, prediction of charmonia profile will not be restricted on the charmonium family members, which determine by $0 \leq L \leq 2$, the investigation involved the higher charmonia that are determined by $2 \leq L \leq 4$. It worth to be mentioned, most of lower charmonia masses have experimentally observed and published by Particle Data Group (PDG), however the other masses of higher states doesn't observed yet. From this perspective, our work will added an incremental contribution by helping PDG to choose the suitable range of energy to observe these states experimentally.

### 4.1 Charmonium wavefunctions

Intuitively in our case, as we use Schrödinger equation and non-relativistic framework, the normalized radial wavefunctions are independent on the selected potential model or even the chosen numerical schemes. So, the illustration of the selected wavefunctions relies upon these results that are acquired form matrix method and the chosen potential model as represented in figures 1 and 2.

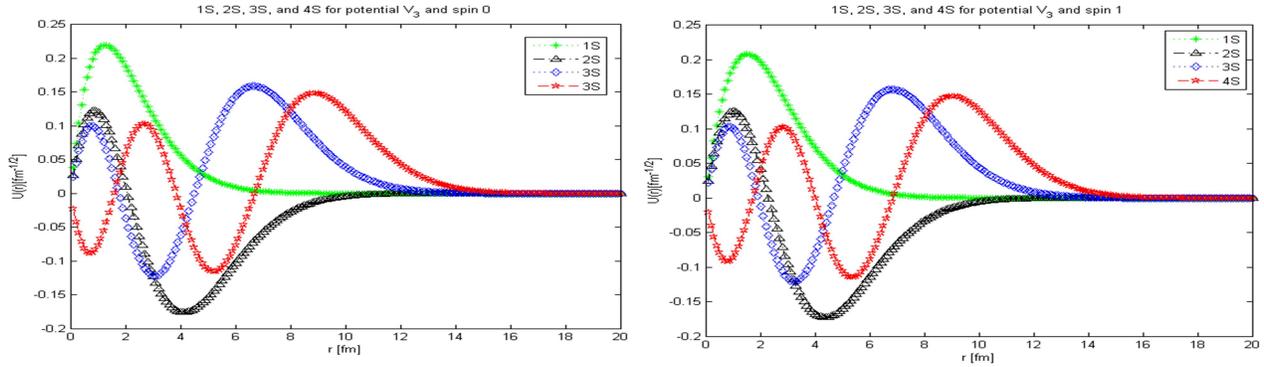

**Figure 1** S-states reduced radial wavefunctions calculated within [$V_3$] potential models for both zero and one spins. We found the S-states radial wavefunctions are spin-undependable.

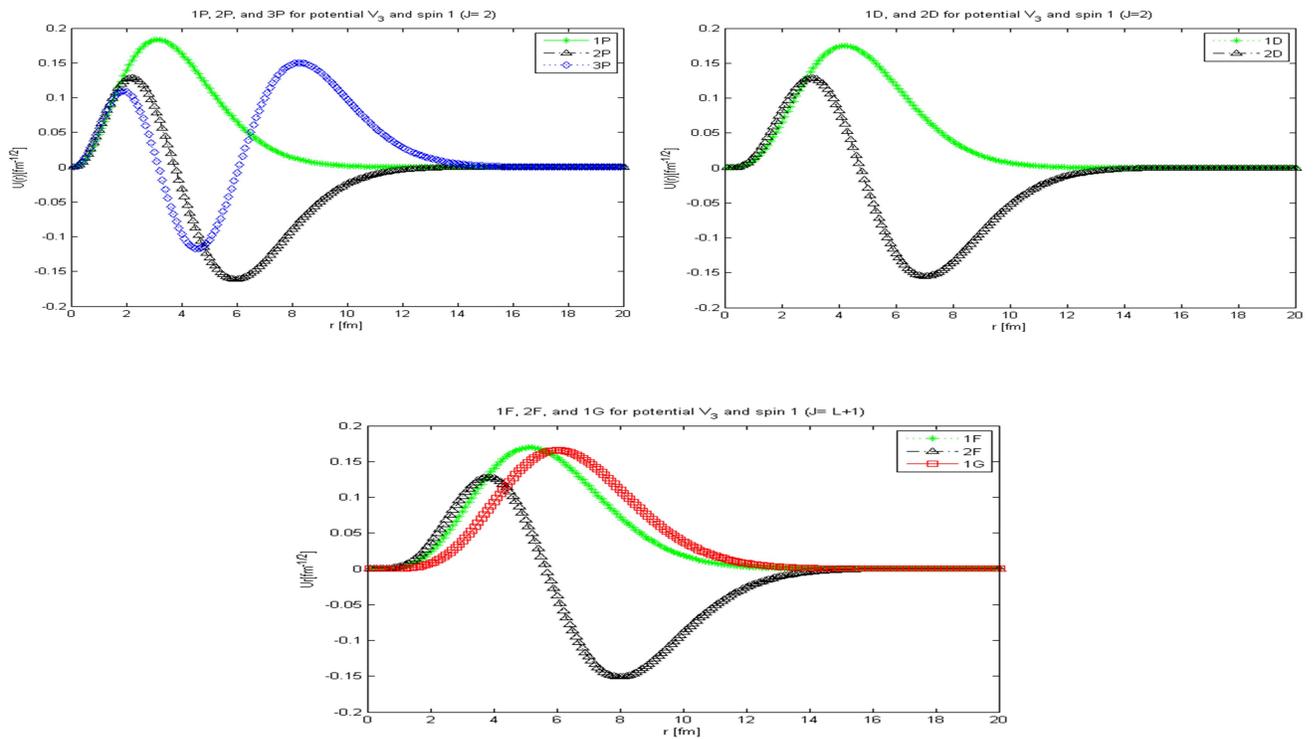

**Figure 2** P-states, D-states, and G-state reduced radial wavefunctions calculated using [$V_3$] potential model for spin one.

## 4.2 Theoretical Calculated Results

Theoretical results for charmonia multiplets are calculated via self-developed proposed codes using MATLAB programming language, and some of them involved some built-in functions of MATLAB program. Additionally, theoretical masses of charmed-Quarkonium are computationally evaluated based on using phenomenological parameters derived from Quantum Chromodynamics (QCD) theory and non-relativistic quark model to fit the corresponding spectra of charmonium theoretical profile, see the fitting parameters for charmoned-Quarkonium in table 1.

**Table 1**: The parameters used in different potentials to fit the result masses of Charmonium states in GeV according to QCD theory.

| Parameters / Potential | Theoretical (NR) ($V_2$) Potential | Theoretical (NR) ($V_3$) Potential |
|---|---|---|
| $m_c = m_{\underline{c}}$ | 1.4399 | 1.4619 |
| $\alpha_s$ | 0.4827 | 0.4942 |
| b | 0.1488 | 0.1446 |
| σ | 1.2819 | 1.1412 |

As mentioned in table 1, we involved an additional potential models denoted by $V_2$. Actually, it is not a new proposed model that we have used in our investigation. It is the same model that we have explained previously in eq.(2.1) without including spin-orbit terms and tensor part to be:

$$V_2(r) = \frac{-4}{3}\frac{\alpha_s}{r} + br + \frac{4\alpha_s}{3m_Q m_{\underline{Q}}}\frac{8\pi}{3}\delta(\vec{r})(\vec{S}_1 \cdot \vec{S}_2)$$

We proposed this expression, for computational simplicity, in conversion test to reduce the run time, as we observed that the yielded results are slightly changed by including these terms to the potential expression. Additionally in this part of the investigation, achieving the convergence of each scheme is the essential aim of this test and also comparing the corresponding iteration number of each method at which the convergence accomplished. Ultimately, the theoretical profile of charmonium multiplets relies upon using potential three, which involved all of these terms.

Convergence investigation will start by illustrating S-states stability curves versus the iteration number for each numerical scheme. Some of these states are singlet states, which obey to $0^{-+}$ type such as $\eta_c(1S)$, $\eta_c(2S)$, and $\eta_c(3S)$. These states possess zero-spin and below the $D\underline{D}$ threshold with an exceptional case for $\eta_c(3S)$. One of them has not observed experimentally yet, which is $\eta_c(4S)$, here it is predicted theoretically. The convergence investigations for $0^{-+}$ states of charmonium are illustrating in table 2 and figure 3 for tri-diagonal matrix method, while table 3 and figure 4 illustrates these results for matrix Numerov's method.

**Table 2**: The stability results of S-States of Charmonium for spin zero using matrix method.

| States | Iteration Number N where at stability occurs | Theoretical mass in GeV |
|---|---|---|
| $\eta_c(1S)$ | 54 | 2.9881 |
| $\eta'_c(2S)$ | 116 | 3.6160 |
| $\eta_c(3S)$. | 140 | 4.0334 |
| $\eta_c(4S)$ | 144 | 4.3809 |

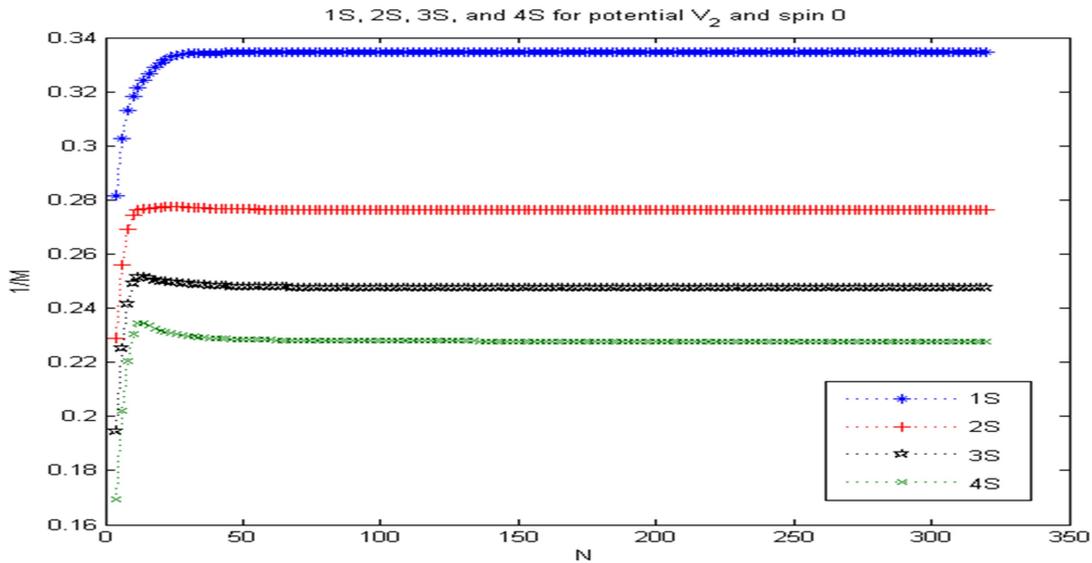

**Figure 3**: S-states stability curves for Charmonium using potential $V_2$ and matrix method.

**Table 3**: The stability results of S-States of Charmonium for spin zero using matrix Numerov's method.

| States | Iteration Number N where at stability occurs | Theoretical mass in GeV |
|---|---|---|
| $\eta_c(1S)$ | 194 | 2.9915 |
| $\eta'_c(2S)$ | 122 | 3.6191 |
| $\eta_c(3S)$. | 168 | 4.0371 |
| $\eta_c(4S)$ | 132 | 4.3854 |

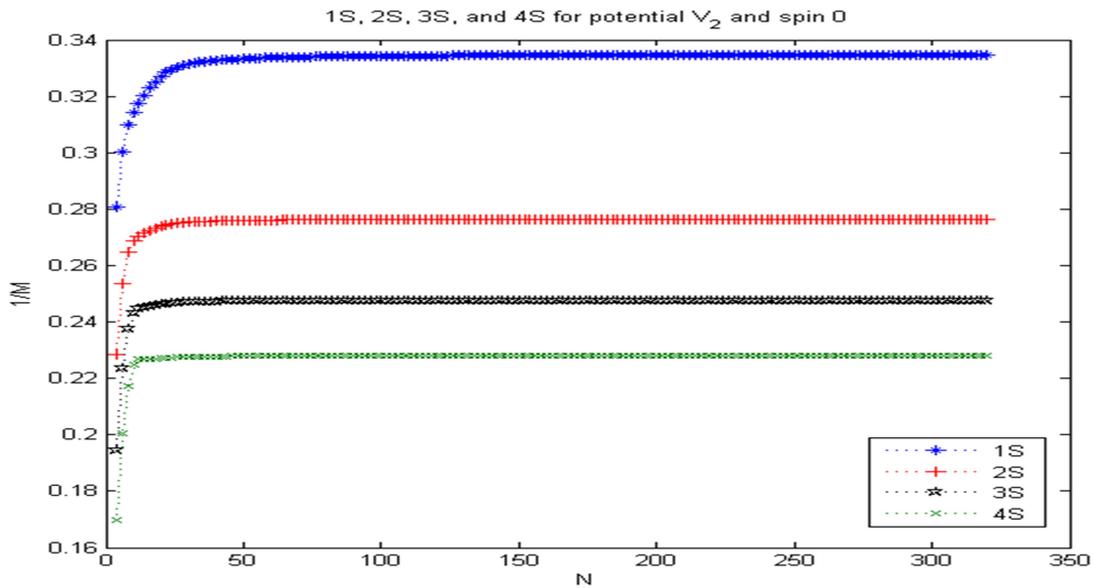

**Figure 4:** S-states stability curves, with spin zero, for Charmonium using potential $V_2$ and using Numerov's matrix method.

The previous illustration shows that both numerical schemes are convergent for $0^{-+}$ states, in spite of matrix Numerov's method observed to be a computationally expensive for these states as it needs more iteration procedures to achieve the convergence. It seems a pre-judge to say that matrix method is preferable computationally. Our investigation will not be restricted to study convergence versus iteration number only.

The next step of our numerical analysis is to illustrate the stability curves of triplet states of S-states that possess spin one. These states called $1^{--}$ states, all of them have observed experimentally. It worth to be mentioned that there is an argument existing between experimentalists about the last discovered state of $1^{--}$ group, if it is a ψ(4S) or a hybrid meson ($\pi^+ \pi^-$ - J/ψ). The corresponding results of $1^{--}$ states for both methods are representing in table 4 with illustration for convergence in figures 5 and 6 as following:

**Table 4**: The stability results of S-States of Charmonium for spin one using Matrix method and Numerov's matrix method for potential $V_2$.

| States | Iteration Number N where at stability occurs | Theoretical mass in GeV | Iteration Number N where at stability occurs | Theoretical mass in GeV |
|---|---|---|---|---|
| N.M | Matrix method | | Numerov's matrix method | |
| ψ(1S) | 144 | 3.0955 | 194 | 3.0959 |
| ψ'(2S) | 84 | 3.6652 | 124 | 3.6694 |
| ψ(3S). | 160 | 4.0699 | 168 | 4.0706 |
| ψ(4S) | 76 | 4.4115 | 134 | 4.4125 |

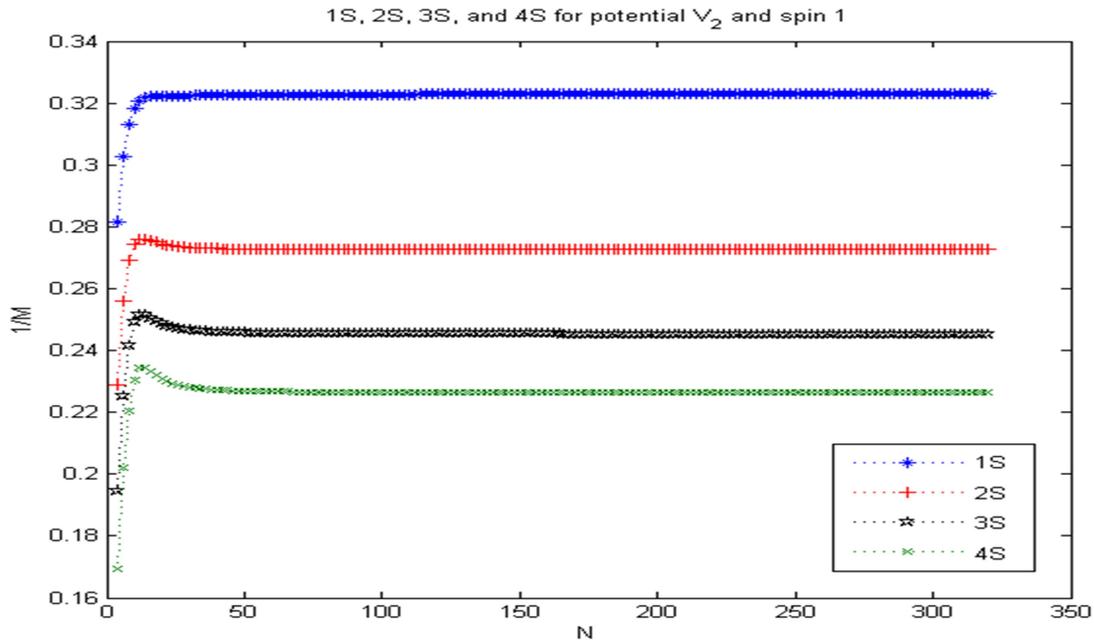

**Figure 5:** S-states stability curves, with spin one, for Charmonium using potential $V_2$ and using matrix method.

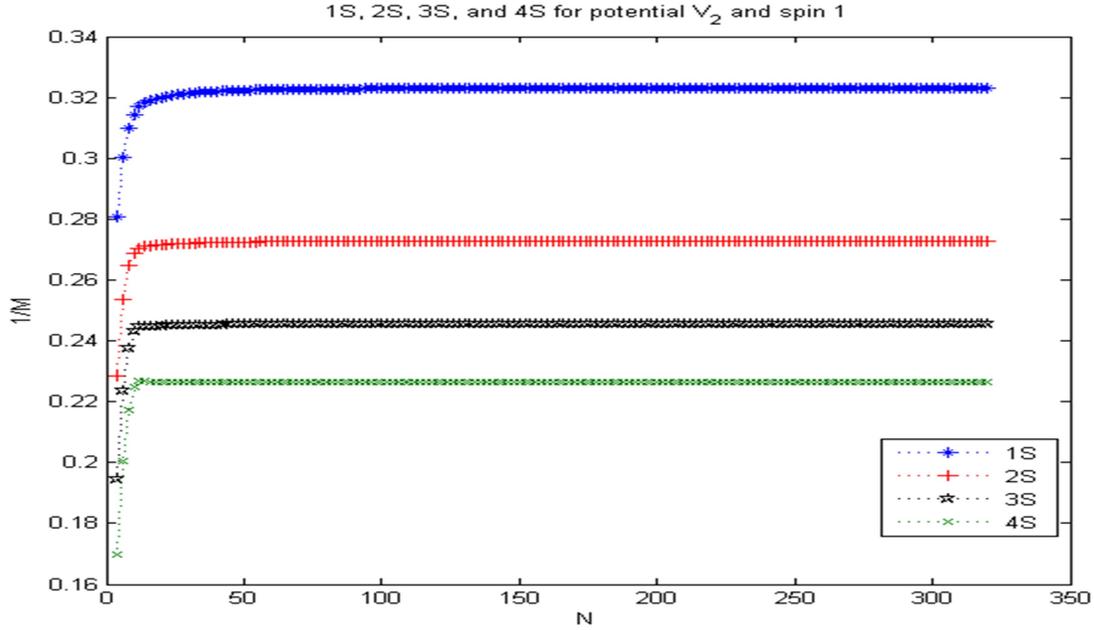

**Figure 6:** S-states stability curves, with spin one, for Charmonium using potential $V_2$ and using Numerov's matrix method.

The next group in our convergence study is for P-states of charmonium, which involved $0^{++}$, $1^{++}$, $1^{+-}$, $2^{++}$ classification. Most of observed P-states are triplet type except $h_c(1P)$ state is singlet one. It worth to be mentioned, all of 1P-states are below $D\underline{D}$ threshold. A comparison between the stability results of both methods for zero-spin P-states is displayed through table 5, while the same comparison for one-spin p-states is represented via table 6 respectively. Starting from P-states, we will represent the yielded results of convergence test versus iteration number only using tables for simplicity.

**Table 5**: The stability results of P-States of Charmonium for spin zero using Matrix method and Numerov's matrix method for potential $V_2$.

| States | Iteration Number N where at stability occurs | Theoretical mass in GeV | Iteration Number N where at stability occurs | Theoretical mass in GeV |
|---|---|---|---|---|
| N.M | Matrix method | | Numerov's matrix method | |
| $h_c(1P)$ | 86 | 3.4989 | 40 | 3.5835 |
| $h_c(2P)$ | 122 | 3.9230 | 114 | 3.9884 |
| $h_c(3P)$. | 84 | 4. 2766 | 62 | 4.3326 |

Table 6: The stability results of P-States of Charmonium for spin one using Matrix method and Numerov's matrix method for potential $V_2$.

| States | Iteration Number N where at stability occurs | Theoretical mass in GeV | Iteration Number N where at stability occurs | Theoretical mass in GeV |
|---|---|---|---|---|
| N.M | Matrix method | | Numerov's matrix method | |
| $\chi_{c2}(1P)$ | 80 | 3.5054 | 44 | 3.5866 |
| $\chi_{c2}(2P)$ | 88 | 3.9307 | 48 | 3.9926 |
| $\chi_{c2}(3P)$. | 84 | 4. 2847 | 100 | 4.3374 |

It was found that both schemes achieved convergence with a computationally cheap iteration number for P-states, but in opposite to the previous case for S-states it was observed the matrix Numerov's method is favored in that case as convergence occurred for small iteration numbers with some exceptions for some P-states like $\chi_{c2}(3P)$ and $h_c(2P)$.

On the other hand, it was observed that the convergence needs approximately the same range of iteration numbers to be occurred using both used numerical schemes for zero and one-spin D-states as it displayed through tables 7 and 8.

Table 7: The stability results of D-States of Charmonium for spin zero using Matrix method and Numerov's matrix method for potential $V_2$.

| States | Iteration Number N where at stability occurs | Theoretical mass in GeV | Iteration Number N where at stability occurs | Theoretical mass in GeV |
|---|---|---|---|---|
| N.M | Matrix method | | Numerov's matrix method | |
| $\psi_{c3}(1D)$ | 52 | 3.7805 | 40 | 3.5866 |
| $\psi_{c3}(2D)$ | 54 | 3.1480 | 48 | 3.9926 |

Table 8: The stability results of D-States of Charmonium for spin one using matrix method.

| States | Iteration Number N where at stability occurs | Theoretical mass in GeV | Iteration Number N where at stability occurs | Theoretical mass in GeV |
|---|---|---|---|---|
| N.M | Matrix method | | Numerov's matrix method | |
| $\psi_{c3}(1D)$ | 46 | 3.7809 | 44 | 3.5866 |
| $\psi_{c3}(2D)$ | 58 | 3.1480 | 46 | 3.9926 |

The previous investigation emphasized that both numerical schemes are reliable to study the spectra of heavy mesons as they are converge and do not blowup to infinity. Additionally, both of them are not expensive computationally as they need small number of iteration processes to meet convergence. In spite of the tri-diagonal matrix method was favored in most cases particularly in lower states of charmonia; the tri-diagonal Numerov's method was not the worst in term of comparison. This method is computationally suitable to be used and it was preferable, according to our investigation, through the higher charmonia states around DD threshold.

Our convergence investigation will not be restricted in studying the stability versus the iteration number; it is involving screening the convergence of our schemes versus the phenomenological radial distance between the charm-anticharm pair. Similarly, in this examination, the inverse of the corresponding charmonium multiplets mass will be investigated versus the radial distance in Fermi units with using a fixed iteration number N=200 and potential $V_2$ to find the optimal radial distance at which the convergence occurred to be used in finding the charmonia mass profile using the sophisticated potential model $V_3$ that we have discussed previously.

According to the results of stability versus the radial distance, it was observed that the phenomenological radial distance for the corresponding charmonium states at which the convergence occurred are approximately the same for the both numerical schemes under the study. This observation increases the credibility of the yielded results of these methods, as the predicted radial distance of each state relies only upon the phenomenological dynamics of the heavy quark-antiquark pair. Such dynamics depend only on the ingredients of QCD.

The illustration of this investigation results will start by the corresponding data for the both spins of S-states as displayed in table 9 and 10 associated with illustration in figures 7, 8, 9 and 10.

**Table 3.9:** The stability results of S-States of Charmonium versus radial distance in Fermi for spin zero using Numerov's matrix and Matrix methods.

| States | The redial distance at which stability occurs using Numerov's matrix method [fm] | The redial distance at which stability occurs using Matrix method [fm] |
|---|---|---|
| $\eta_c(1S)$ | 8 | 8 |
| $\eta'_c(2S)$ | 10 | 10 |
| $\eta_c(3S)$. | 14 | 12 |
| $\eta_c(4S)$ | 18 | 18 |

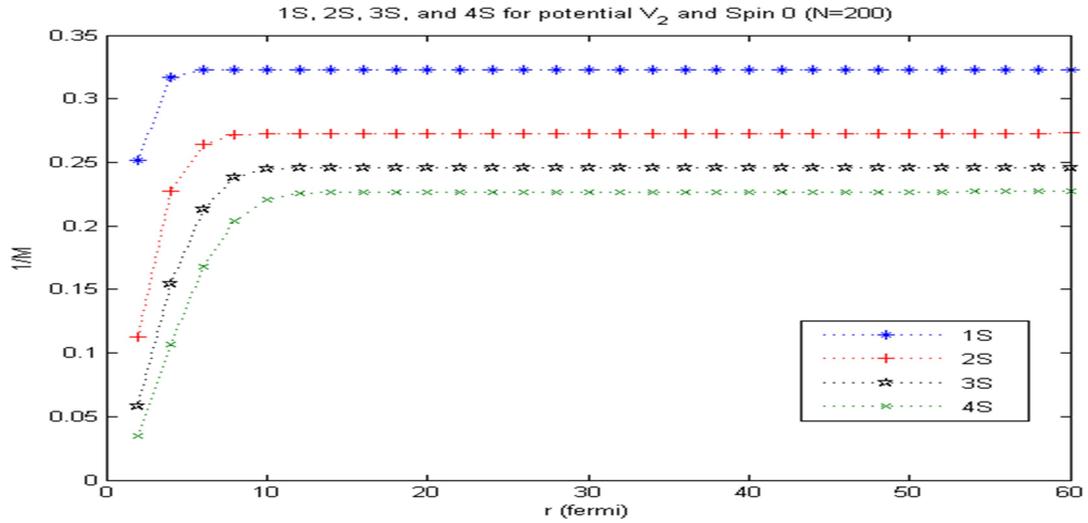

**Figure 7:** S-states stability curves, with spin zero, for Charmonium using potential $V_2$ and using Matrix method.

**Table 10:** The stability results of S-States of Charmonium versus radial distance in Fermi for spin one using Numerov's matrix and Matrix methods.

| States | The redial distance at which stability occurs using Numerov's matrix method [fm] | The redial distance at which stability occurs using Matrix method [fm] |
|---|---|---|
| $J/\psi_c(1S)$ | 8 | 8 |
| $\psi'_c(2S)$ | 10 | 10 |
| $\psi_c(3S)$. | 14 | 12 |
| $\psi_c(4S)$ | 18 | 18 |

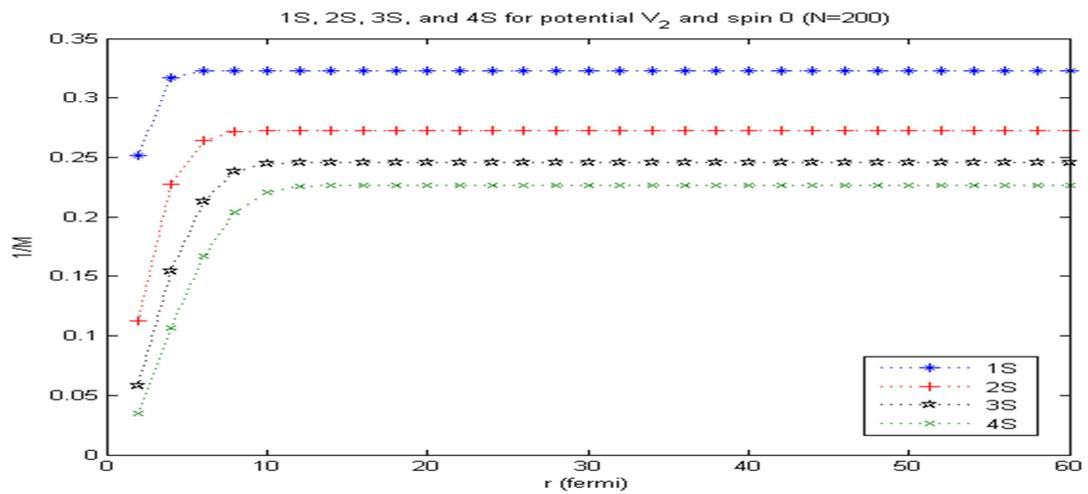

**Figure 8:** S-states stability curves, with spin zero, for Charmonium using potential $V_2$ and using matrix Numerov's method.

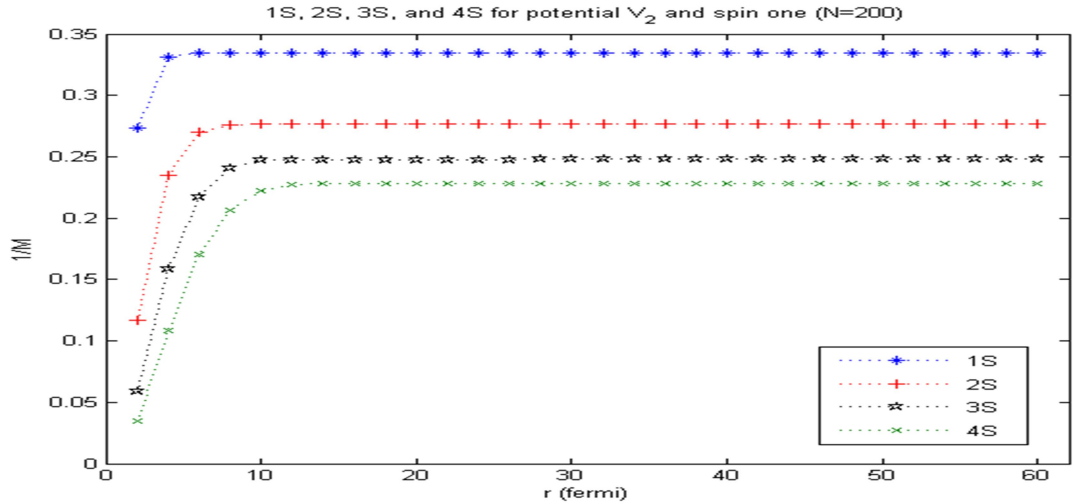

**Figure 9:** S-states stability curves, with spin one, for Charmonium using potential $V_2$ and using Matrix method.

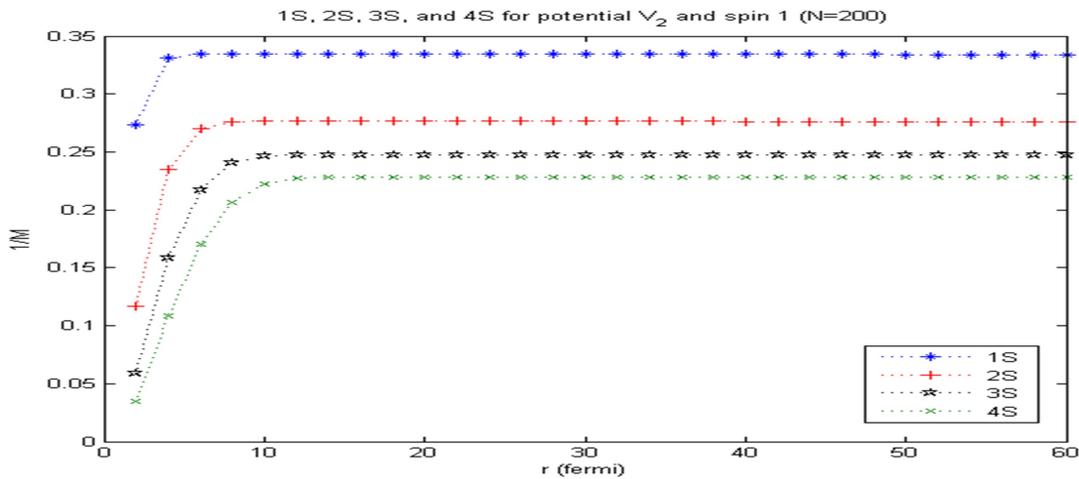

**Figure 10:** S-states stability curves, with spin one, for Charmonium using potential $V_2$ and using Numerov's matrix method.

The previous illustrations for S-states and the rest of the yielded results for both P- and D-states emphasized that the convergence curves versus the radial distance are completely phenomenological without dependency on the used numerical scheme or even the spin as shown in the illustration of P-states results for both spins that represented in tables 11 and 12 and figures 11 and 12.

**Table 11:** The stability results of P-States of Charmonium versus radial distance in Fermi for spin zero using Numerov's matrix and Matrix methods.

| States | The redial distance at which stability occurs using Numerov's matrix method [fm] | The redial distance at which stability occurs using Matrix method [fm] |
|---|---|---|
| $h_c(1P)$ | 10 | 10 |
| $h_c(2P)$ | 12 | 12 |
| $h_c(3P)$. | 14 | 14 |

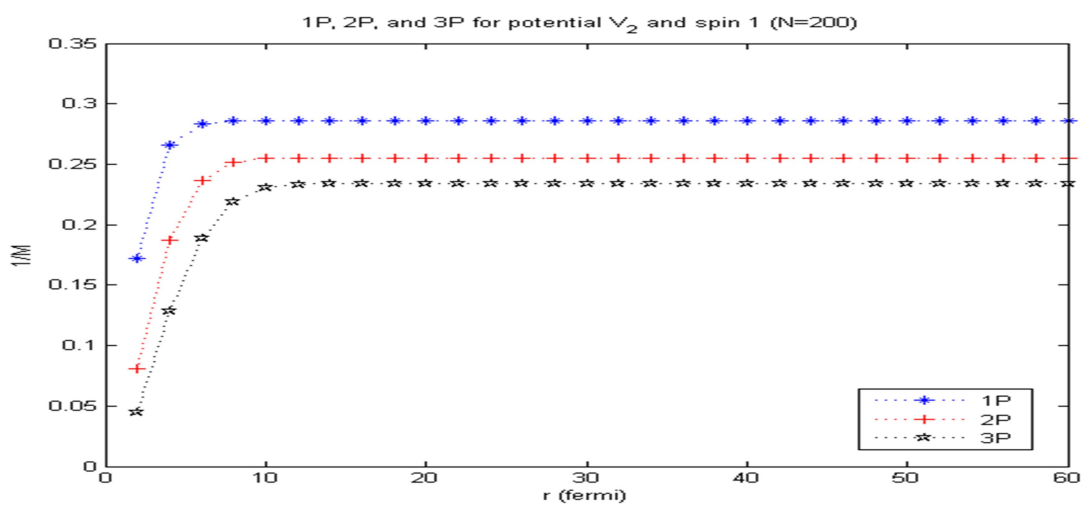

**Figure 11:** P-states stability curves, with spin zero, for Charmonium using potential $V_2$ and using Matrix method. The same results are obtained for Matrix Numerov's method.

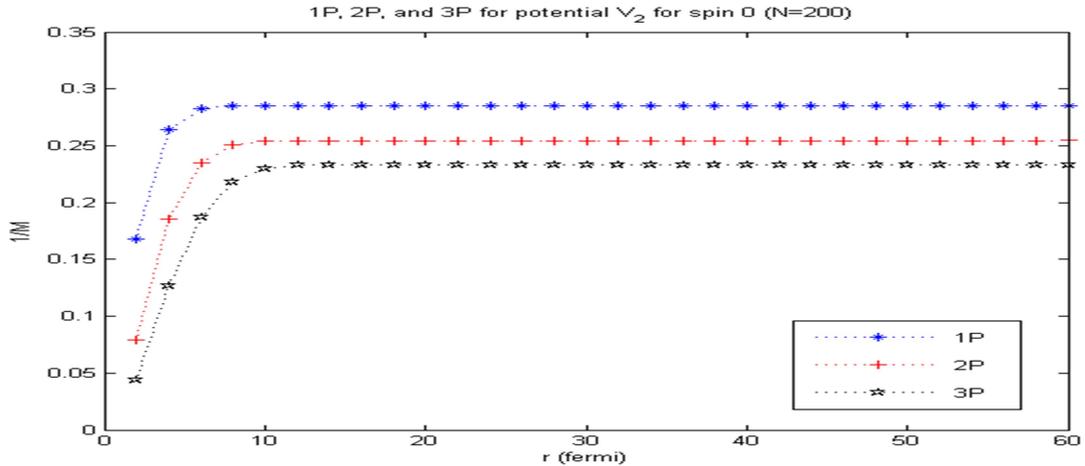

**Figure 12:** P-states stability curves, with spin one, for Charmonium using potential $V_2$ and using Matrix method. Similar curves are obtained for Matrix Numerov's method.

**Table 12:** The stability results of P-States of Charmonium versus radial distance in Fermi for spin one using Numerov's matrix and Matrix methods.

| States | The redial distance at which stability occurs using Numerov's matrix method [fm] | The redial distance at which stability occurs using Matrix method [fm] |
|---|---|---|
| $\chi_{c2}(1P)$ | 8 | 8 |
| $\chi_{c2}(2P)$ | 12 | 12 |
| $\chi_{c2}(3P)$. | 14 | 14 |

The same observation is emphasized from the results of D-states as the convergence is occurred at the same phenomenological radial distance with independency on the numerical recipes of spins as displayed in table 13 below.

**Table 13:** The stability results of D-States of Charmonium versus radial distance in Fermi for zero and one spins using matrix Numerov's and Matrix methods.

| States | The redial distance at which stability occurs using Numerov's matrix method [fm] | The redial distance at which stability occurs using Matrix method [fm] |
|---|---|---|
| $\psi_{c2}(1D)$ & $\psi_{c3}(1D)$ | 10 | 10 |
| $\psi_{c2}(2D)$ & $\psi_{c3}(2D)$ | 12 | 12 |

From the previous convergence investigations, it was observed that the optimal iteration number to be used by both numerical schemes for all charmonia is N=200. Although the convergence occurred for some charmonia multiplets near 10 Fermi, it is preferable to choose the optimal radial distance to be 20 Fermi for two reasons that will be elaborated respectively.

The first reason for choosing this is to avoid the phenomenon of the color confinement that might be occurred among low radial distances. While the second is to be suitable for all multiplets of charmonia, as some of them need radial distances around 18 Fermi to achieve the convergence as it was discussed previously. Ultimately, both factors, N=200 as iteration sequences and r= 20 Fermi, will be considered in the final comparison to find the favored numerical scheme to study heavy charmed Quarkonium.

The last investigation relies upon comparing the theoretical profiles of charmonium yielded by the two numerical schemes understudy with the experimental mass-spectrum provided by PDG. This investigation depends on the minimization of the $\chi^2$ function, which as previously discussed, to find the optimal scheme that achieves a high endorsement with experiments.

The theoretical mass profiles involve 28 multiplets of charmed Quarkonium evaluated by our two tridiagonal matrix methods with two potential models, which the second one involves all the hyperfine spin-spin and spin-forces interactions. Intuitively, the second potential is found to be preferable in terms of comparison with experiments. Additionally, it was observed that the desired optimization is occurred using the tridiagonal matrix method, which achieved the optimal minimization for the $\chi^2$ function to be 0.0001 using the V3 potential model. On the other hand, the investigation emphasized that Matrix Numerov's method is suitable also for studying the spectroscopy of heavy mesons as the endorsement between the theoretical profile and experiments was acceptable as its associated minimization was 0.0042 via the same potential model. The theoretical profiles and experimental results for charmonia are represented in the table 14.

**Table 14**: Experimental and theoretical spectra of Charmonium [$c\underline{c}$] states, in $GeV$, for various computational grid number. The considered $c\underline{c}$ radius is equal 20 Fermi.

| State | name | Exp. Mass [PDG 2010] | Theoretical masses for grid (200 × 200) | | Theoretical masses for grid (200 × 200) | |
|---|---|---|---|---|---|---|
| Potential model | | | $V_2$ | $V_3$ | $V_2$ | $V_3$ |
| Numerical scheme | | | Matrix method [35] | | Numerov's matrix method | |
| $1^3S_1$ | J/ψ | 3.9687 ± 0.0004 | 3.0955 | 3.1129 | 3.0959 | 3.1134 |
| $1^1S_0$ | $\eta_c(1S)$ | 2.9792 ± 0.0013 | 2.9881 | 3.0171 | 2.9890 | 3.0179 |
| $2^3S_1$ | $\psi(2S)$ | 3.68609 ± 0.0004 | 3.6652 | 3.6779 | 3.6694 | 3.6785 |
| $2^1S_0$ | $\dot{\eta}(2S)$ | 3.637 ± 0.0004 | 3.6171 | 3.6373 | 3.6179 | 3.6382 |
| $3^3S_1$ | $\psi(3S)$ | 4.039 ± 0.001 | 4.0699 | 4.0757 | 4.0706 | 4.0765 |
| $3^1S_0$ | $\eta_c(3S)$ | | 4.0352 | 4.0473 | 4.0362 | 4.0482 |
| $4^3S_1$ | $\psi(4S)$ | 4.421 ± 0.0004 | 4.4115 | 4.4107 | 4.4125 | 4.4117 |
| $4^1S_0$ | $\eta_c(4S)$ | | 4.3835 | 4.3880 | 4.3846 | 4.3891 |
| $1^3P_2$ | $\chi_2(1P)$ | 3.55620 ± 0.00009 | 3.5054 | 3.5452 | 3.9926 | 3.6197 |
| $1^3P_1$ | $\chi_2(1P)$ | 3.51066 ± 0.00007 | 3.5054 | 3.5077 | 3.9926 | 3.6005 |
| $1^3P_0$ | $\chi_0(1P)$ | 3.41475 ± 0.00031 | 3.5054 | 3.3996 | 3.9926 | 3.4936 |
| $1^1P_1$ | $h_c(1P)$ | 3.52541 ± 0.00016 | 3.4989 | 3.5175 | 3.9884 | 3.6050 |
| $2^3P_2$ | $\chi_2(2P)$ | | 3.9307 | 3.9611 | 3.9926 | 3.961 |
| $2^3P_1$ | $\chi_2(2P)$ | | 3.9307 | 3.9256 | 3.9926 | 3.9964 |
| $2^3P_0$ | $\chi_0(2P)$ | | 3.9307 | 3.8470 | 3.9926 | 3.8840 |
| $2^1P_1$ | $h_c(2P)$ | | 3.9230 | 3.9341 | 39884 | 4.0014 |
| $3^3P_2$ | $\chi_2(3P)$ | | 4.2847 | 4.3073 | 4.3374 | 4.3547 |
| $3^3P_1$ | $\chi_2(3P)$ | | 4.2847 | 4.2727 | 4.3374 | 4.3329 |
| $3^3P_0$ | $\chi_0(3P)$ | | 4.2847 | 4.2083 | 4.3374 | 4.2679 |
| $3^1P_1$ | $h_c(3P)$ | | 4.2766 | 4.2807 | 4.3362 | 4.3380 |
| $1^3D_3$ | $\psi_3(1D)$ | | 3.7809 | 3.7991 | 3.8952 | 3.9117 |
| $1^3D_2$ | $\psi_2(1D)$ | | 3.7809 | 3.7977 | 3.8952 | 3.9166 |
| $1^3D_1$ | $\psi(1D)$ | $3.77292 \pm 0.00035$ | 3.7809 | 3.7829 | 3.8952 | 3.9120 |
| $1^1D_2$ | $\psi_{c2}(1D)$ | | 3.7805 | 3.7959 | 3.8951 | 3.9136 |
| $2^3D_3$ | $\psi_3(2D)$ | | 4.1488 | 4.1611 | 4.2450 | 4.2550 |
| $2^3D_2$ | $\psi_2(2D)$ | | 4.1488 | 4.1568 | 4.2450 | 4.2573 |
| $2^3D_1$ | $\psi(2D)$ | $4.153 \pm 0.0003$ | 4.1488 | 4.1400 | 4.2450 | 4.2507 |
| $2^1D_2$ | $\psi_{c2}(2D)$ | | 4.1480 | 4.1557 | 4.2448 | 4.2551 |
| | $\chi^2$ | | 0.0011 | $0.0000738 \approx 0.0001$ | 0.0054 | 0.0042 |

## 5. Conclusion

In our approach, two different tridiagonalization schemes have been used to develop two numerical methods to reduce the solution of one dimension radial Schrödinger equation into an eigenvalue problem to study the spectroscopy of charmed-Quarkonium using a potential model derived QCD, which involves the phenomenological contributions of heavy quark-antiquark interactions according to the ingredients of both QCD theory and non-relativistic quark model.

Our two numerical methods obeyed numerical analysis investigations such as the convergence test versus the interaction sequences and the phenomenological radial distance between the heavy charmed-quarks pair. Investigations emphasized that both of them are reliable to study the spectroscopy of charmonium and heavy mesons, as they converge and do not blow up into infinity. Additionally, they are not expensive computationally as they allow the convergence to be occurred by using a computational grid as low as 100 x 100 in most cases and sometimes the desired minimization as low as 50 x 50 in the grid size achieved, particularly for the low states of charmonium before the $D\underline{D}$ threshold.

It was observed that our numerical recipes provide a very good approximation for handling the members of the heavy mesonic family. Both of them exhibit a satisfactory agreement with experiments, as a persuaded minimization for $\chi^2$ values has found for each of them. But the desired minimization occurred for the matrix method, which based on using the finite difference algorithm in the tridiagonal matrix form without losing the substantial minimization corresponding to the grid size. Ultimately, the description of the heavy mesonic family is better with using the tri-diagonal matrix method.